\documentclass[reprint,twocolumn,superscriptaddress,secnumarabic,amssymb,nobibnotes,aps,prb]{revtex4-2}

\usepackage{times}
\usepackage{graphicx}
\usepackage{dcolumn}
\usepackage{bm}
\usepackage{amsmath}
\usepackage[colorlinks=true, allcolors=blue]{hyperref}
\usepackage{ulem}
\usepackage{chngcntr} 
\usepackage{enumitem}
\usepackage[version=3]{mhchem}  
\usepackage{lineno}

\newcommand{\fmo}{\ce{Fe2(MoO4)3}}

\begin{document}
\title{Impact of Oxygen Vibrations on Raman Intensity Variation in the Catalyst \fmo}

\author{Young-Joon Song}
\author{Roser Valent\'{i}}
\affiliation{Institut f\"ur Theoretische Physik, Goethe-Universit\"at Frankfurt, 
Max-von-Laue-Stra\ss e 1, 60438 Frankfurt am Main, Germany}
\date{\today}

\begin{abstract}
Iron molybdate (\fmo) is a widely used commercial catalyst for oxidative dehydrogenation. Recently, the possibility that bulk oxygen atoms participate in catalytic reactions has been proposed based on the experimentally observed significant reduction in Raman intensity during the catalytic process, which implies the formation of oxygen defects. In this work, we performed density functional theory (DFT) calculations to elucidate the microscopic mechanism of the experimentally observed Raman intensity variation. Our phonon analysis reveals that oxygen-dominated vibrational modes, with a small contribution from Mo, occur near 782cm$^{-1}$-- the same frequency region where the Raman intensity reduction was measured. To make the calculations computationally feasible for this large system, we introduced an effective frozen-phonon approach to indirectly account for the effects of a very small concentration of defects on the Raman intensity. Our results suggest that oxygen vibrations are primarily responsible for the decrease in the calculated Raman intensity. Moreover, structural relaxation of \fmo\ containing an oxygen vacancy indicates that oxygen diffusion from the bulk to the surface may occur very rapidly, such that the local symmetry remains effectively unchanged. This interpretation is in line with the absence of measurable peak shifts or broadening in the experimental Raman spectra.
\end{abstract}

\maketitle
\section{Introduction}
Iron molybdate (\fmo) has attracted significant attention owing to its intriguing magnetic, electrochemical, and catalytic properties. The system exhibits semiconducting behavior and undergoes an L-type ferrimagnetic transition below 12 K. \cite{Mari1985,battle_study_1982,forzatti_electrical_1984,parveen_synthesis_2020,tiwari_observation_2022, tiwari_modulation_2024} Remarkably, the application of an external magnetic field induces exotic multiferroic phases at even lower temperatures. \cite{tiwari_observation_2022, tiwari_modulation_2024}
As an electrode material, \fmo~is capable of accommodating alkali-ion (e.g., Li and Na) intercalation through robust edge-shared polyhedral units, which provide large interstitial voids and confer high structural stability. This structural flexibility underlies its favorable cycling performance and reversibility. \cite{manthiram_lithium_1987,sun_nasicon-type_2012,kimura_modifications_2024,zhou_influence_2016}

In addition, \fmo~is widely utilized as an industrial catalyst for methanol-to-formaldehyde conversion via oxidative dehydrogenation (ODH). \cite{adkins_oxidation_1931,hess_nanostructured_2011,gaur_operando_2019,bowker_ensemble_2024,schumacher_unraveling_2023} 
It has been reported that dense Mo concentration at the surface is required to achieve higher selectivity to formaldehyde, whereas selectivity decreases in iron-rich regions. However, iron sites are also necessary to fulfill the oxygen demand. \cite{gaur_operando_2019,bowker_ensemble_2024,schumacher_unraveling_2023}
Despite decades of study, the precise catalytic mechanism remains to be elucidated. Recent operando Raman and impedance spectroscopy experiments on the ODH process by Schumacher {\it et al.} \cite{schumacher_unraveling_2023} proposed that the bulk region of \fmo~may act as an oxygen reservoir, in contrast to the conventional view that catalytic activity originates primarily at the surface or subsurface. Their work revealed a pronounced reduction in Raman intensity at around 782 cm$^{-1}$ during catalysis at 500 \textdegree{}C with increasing conversion, suggesting the formation of oxygen-related defects due to significant oxygen transport from the bulk to the surface of the catalyst. However, the microscopic mechanism of this reduction remains unresolved.

Here, we address the role of oxygen vibrations in the observed Raman intensity reduction through density functional theory calculations (DFT). We first compare the low-temperature monoclinic and high-temperature orthorhombic crystal structures of \fmo, which are found to be nearly identical, leading to very similar phononic structures. Our phonon analysis shows that asymmetric \ce{MoO4} stretching modes dominate at around 789 cm$^{-1}$, with minor Mo vibrations, whereas symmetric stretching modes occur near 972 cm$^{-1}$ with negligible Mo contribution. These results lead to calculated Raman intensities in good agreement with experimental observations. 
Finally, by employing an effective frozen-phonon approach, in which selected phonon modes are manually frozen to distinguish each phonon's contribution, we suggest that oxygen phonon modes are the key factor responsible for the observed reduction in Raman intensity.
To support the validity of our approach, we also analyze the role of local structural distortions on the Raman response when an oxygen vacancy is created in the structure.
Furthermore, our structural relaxation shows that when an oxygen vacancy is introduced, a pentagonal \ce{FeO5} unit remains stable, suggesting that iron sites may play an active role in facilitating oxygen diffusion during the catalytic process.

\section{Calculation methods}
Density functional theory-based \textit{ab initio} calculations were carried out to investigate the electronic and phononic structures of \fmo~using the projector augmented wave (PAW) method \cite{paw}, implemented in the Vienna Ab initio Simulation Package (VASP) \cite{vasp}. The revised generalized-gradient approximation (denoted PBESol throughout the paper) was employed to treat the exchange-correlation functional. \cite{gga,pbesol}
In addition, the onsite Hubbard $U$ of 4 eV was introduced to treat strongly correlated Fe 3$d$ orbitals. \cite{dudarev1998}
The plain-wave basis set size was determined by the energy cut-off of 500 eV, and the Gaussian smearing method with a width of 0.05 eV was utilized. For monoclinic \fmo, the Brillouin zone was sampled by 2$\times$3$\times$2, while a 3$\times$4$\times$4 $k$-mesh is employed for the orthorhombic structure.
In order to obtain the experimentally observed semiconducting behavior with a gap of about 2.7 eV \cite{forzatti_electrical_1984,parveen_synthesis_2020}, spin polarized calculations were carried out, assuming simple ferromagnetic spin alignment leading to a high-spin configuration of Fe$^{3+}$. \cite{battle_study_1982}. 
Otherwise, the electronic structure converges to be metallic within non-spin-polarization in calculations, leading to incorrect phonon and Raman results as discussed further below. \cite{floris_vibrational_2011,miwa_prediction_2018}
Before performing phonon calculations, all structures were fully relaxed until the net forces became smaller than 1 meV/\AA.
The finite displacement method, as implemented in VASP, was used for phonon calculations with the magnitude of a 0.01 \AA~displacement, and then the dynamical matrix was constructed to obtain phonon dispersions and density of states using the phonopy \cite{phonopy-phono3py-JPCM,phonopy-phono3py-JPSJ} package.
For the orthorhombic structure, we additionally adopted a 2$\times$2$\times$2 supercell with a single $k$ point to eliminate the artificial imaginary frequencies that appeared in the phonon calculations performed with a single unit cell. It is worth noting that the phonon modes at the $\Gamma$ point are accurately determined even within a single unit cell, if the system is mechanically stable. Furthermore, we confirmed that the overall phonon spectra obtained from the supercell and single-unit-cell calculations are consistent.
Finally, Raman intensity \cite{raman_dft,vasp_raman_py} was calculated through dielectric tensor results in which the calculated phonon eigenvectors at the $\Gamma$ point were involved.
The figures of the crystal structures were generated with VESTA. \cite{vesta}

\section{Results}
\subsection{Crystal structures: low- and high-temperature polymorphs}\label{sec:crystal_structure}
At ambient conditions, \fmo~crystallizes in a low-temperature monoclinic structure ($P2_1/c$, 8 $f.u.$), as illustrated in Fig. \ref{fig:str_phon_Ram_mono} (a). \cite{chen_crystal_1979}
The relaxed lattice constants are $a = 15.869$ \AA, $b = 9.373$ \AA, and $c = 15.868$ \AA, with the angle $\beta$ of 108.36 \textdegree{}, which is in good agreement with the experimental measurements. \cite{chen_crystal_1979}
All atoms (Fe, Mo, and O) occupy the $4e$ Wyckoff positions, and the corresponding relaxed coordinates are listed in Table \ref{tab:str_info_diff}.
Upon heating, \fmo~undergoes a structural phase transition from the monoclinic phase to an orthorhombic ($Pbcn$) structure at 499 \textdegree{}C. \cite{plyasova_polymorphism_1977,harrison_crystal_1995,sleight_new_1973} (See Fig. \ref{fig:str_phon_Ram_ortho} (a)) 
In both polymorphs, the structure consists of interconnected corner-sharing \ce{FeO6} octahedra and \ce{MoO4} tetrahedra, linked through Fe-O-Mo bonds. 
In the relaxed orthorhombic \fmo, the lattice constants amount to be $a = 12.867 $ \AA, $b = 9.287 $ \AA, and $c = 9.373$ \AA, which is in line with experiment. \cite{harrison_crystal_1995}
The corresponding relaxed coordinates of the nine inequivalent atoms are summarized in Table \ref{tab:str_info_ortho}.

\begin{table}[h]
    \centering
    \caption{Structural information for the relaxed orthorhombic \fmo\ obtained as detailed in 
    the main text.}
    \vspace{1.0 mm}
    \begin{tabular}{ccccc}
    \hline\hline \\
    \vspace{-6.5 mm} \\
    \multicolumn{2}{c}{Space Group} & $a$ (\AA) & $b$ (\AA) & $c$ (\AA) \\
    \hline  
    \multicolumn{2}{c}{No. 60 ($Pbcn$)} & 12.867 & 9.287 & 9.373  \\
    \vspace{-2.5 mm} \\
    %
    Atom & Wyckoff & $x$ & $y$ & $z$ \\
    \hline
    Fe   &  8d  & ~0.8806~  & ~0.7506~  & ~0.5356~  \\
    Mo1  &  4c  &  0.0000  &  0.9765  &  0.2500  \\
    Mo2  &  8d  &  0.8548  &  0.3986  &  0.3788  \\
    O1   &  8d  &  0.0594  &  0.8681  &  0.1190  \\
    O2   &  8d  &  0.3269  &  0.8655  &  0.3029  \\
    O3   &  8d  &  0.0238  &  0.3222  &  0.0788  \\
    O4   &  8d  &  0.7417  &  0.8223  &  0.4893  \\
    O5   &  8d  &  0.9044  &  0.0875  &  0.1711  \\
    O6   &  8d  &  0.3600  &  0.0867  &  0.0945  \\
    \hline\hline    
    \end{tabular} 
    \label{tab:str_info_ortho}
\end{table}

In Ref. \cite{harrison_crystal_1995}, it was noted that only minor differences exist between the two polymorphs in terms of Fe-O-Mo bond angles, Fe-Mo separation, and overall densities.
To further quantify the structural similarity theoretically, we performed a symmetry-mode analysis of the displacive phase transition using the AMPLIMODES program \cite{perez-mato_mode_2010,orobengoa_amplimodes_2009} at the Bilbao Crystallographic Server. 
This analysis identified the transformation matrix for the transition from the orthorhombic ($Pbcn$) to monoclinic ($P2_1/c$) structures, expressed as 
    \[ 
   \begin{bmatrix}
   -1  &  0  & 1   \\
   -1  &  0  & -1   \\
    0  & -1  & 0
   \end{bmatrix}
   ~~~~
   \begin{bmatrix}
    0 \\
    0 \\     
    -\frac{1}{2}
   \end{bmatrix}.
   \]
Differences in atomic positions between the two polymorphs are found to be very small (less than 4$\times$10$^{-3}$ in magnitude), as summarized in Table \ref{tab:str_info_diff}. 
The orthorhombic unit cell contains 68 atoms (4 $f.u.$), which is half the number of atoms in monoclinic \fmo, making calculations more computationally feasible. 
Our DFT results show a total energy difference of only 0.8 meV/f.u. between the two structures, further confirming the structural similarity of the two polymorphs. 
Moreover, since the monoclinic ($C_{2h}$) and orthorhombic ($D_{2h}$) point groups are in a subgroup–supergroup relationship, symmetry-related properties remain compatible.
The structural analysis suggests that the orthorhombic structure serves as an effective and computationally efficient model for the monoclinic phase, which is considerably more demanding for DFT calculations.
In this work, we focus on the high-temperature orthorhombic phase of \fmo, as the pronounced Raman intensity reduction is observed at 500 \textdegree{}C, whereas only negligible changes are detected at 300 \textdegree{}C, where the crystal structure retains the monoclinic space group. \cite{schumacher_unraveling_2023}
For comparison, we also analyzed the monoclinic structure, where the phonon and Raman results are nearly identical to those from the orthorhombic phase.

\subsection{Phonon and Raman properties}
\begin{figure*}[ht]
\includegraphics[width=2.0\columnwidth]{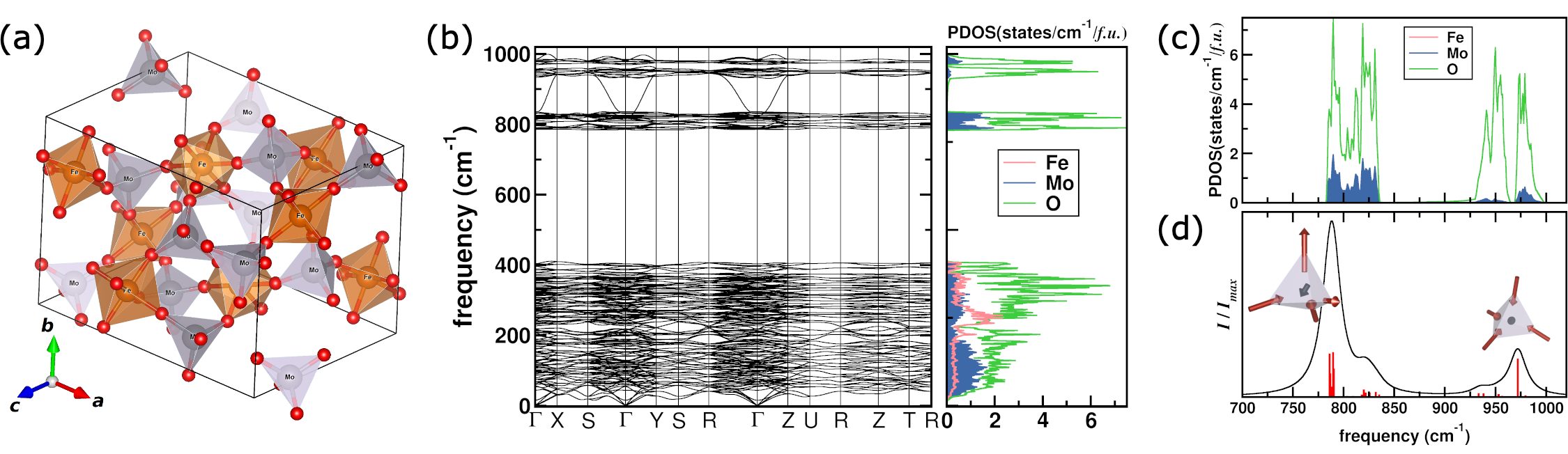} 
\caption{
(a) Crystal structure of high-temperature orthorhombic \fmo. (b) The corresponding phonon dispersions along the high symmetry points and atom-resolved phonon density of states. (c) Enlarged atom-resolved phonon density of states of the high-frequency range. 
(d) Calculated Raman intensity ($I$), normalized to the maximum value ($I_{max}$), depicted with a thin red bar, as a function of frequency for orthorhombic \fmo. 
Lorentzian broadening with a width of 10 cm$^{-1}$ is plotted with a black solid line. 
At 789.68 cm$^{-1}$, asymmetric \ce{MoO4} vibrational modes are dominant, while symmetric ones are found at 972.05 cm$^{-1}$.
A red (black) arrow represents the O(Mo)-atom vibrations at a given frequency.  
}
\label{fig:str_phon_Ram_ortho}
\end{figure*}

In the previous section, we discussed the structural similarity between the two relaxed polymorphs. 
Consistently, our phonon calculations show that their vibrational properties are nearly identical, which can be seen in Fig. \ref{fig:str_phon_Ram_ortho} (b) and Fig. \ref{fig:str_phon_Ram_mono} (b), where the phonon dispersions along the high-symmetry points of the first Brillouin zone and the corresponding atom-resolved phonon density of states are presented for orthorhombic and monoclinic \fmo, respectively.
It should be noted that the presence of the artificial frequencies disappears when a 2$\times$2$\times$2 supercell is utilized, which is confirmed for the orthorhombic structure. 
Importantly, no imaginary modes appear at the $\Gamma$ point in either case, independent of the unit cell size, ensuring that the Raman-active phonons remain well defined, since Raman experiments can only measure excitations in the $\vec{q} \rightarrow 0$ limit. \cite{kitajima_defects_1997}

In both polymorphs, the phonon spectrum separates into two distinct regimes: a low-frequency range (0 to 403 cm$^{-1}$), where all Fe, Mo, and O atoms contribute, and a high-frequency range (780 to 1001 cm$^{-1}$), where the Fe contribution vanishes.
In general, the phonon frequency $w$ is proportional to $\sqrt{\frac{k_{eff}}{\mu}}$, where $k_{eff}$ and $\mu$ denote the effective force constant and the relevant reduced mass, respectively. Therefore, if $k_{eff}$ is comparable, lighter elements are expected to have higher phonon frequencies. In \fmo, the Mo--O bonds ($\sim$ 1.77 \AA) are shorter than the Fe--O bonds ($\sim$ 1.97 \AA), resulting in stronger $p-d$ orbital hybridization within the MoO$_4$ tetrahedra. In addition, the more spatially extended nature of the 4$d$ orbitals compared to the 3$d$ orbitals further enhances this hybridization. Consequently, the Mo-O bonds are much stiffer, leading to larger force constants. This explains why phonon modes involving Mo contributions appear in the high-frequency region.
In particular, the Mo-derived phonon modes appear more in the 780-830 cm$^{-1}$ range, whereas their contribution becomes negligible above 830 cm$^{-1}$, as highlighted in Fig. \ref{fig:str_phon_Ram_ortho} (c), where the high-frequency region of the atom-resolved phonon density of states for orthorhombic \fmo~is enlarged. 
Consequently, Raman features observed between 780 and 830 cm$^{-1}$ can be expected to originate from mixed Mo and O vibrations, while those above this range are primarily associated with O-atom vibrations.
For the monoclinic ($P2_1/c$) structure, Raman-active modes at the $4e$ Wyckoff position are based on the $A_g$ and $B_g$ irreducible representations (irreps). 
In contrast, for the orthorhombic ($Pbcn$) structure, where two additional rotational axes are present, Raman activity arises from the $A_g$, $B_{1g}$, $B_{2g}$, and $B_{3g}$ irreps associated with the $4c$ and $8d$ Wyckoff positions. 
The calculated frequencies of the Raman-active optical phonons at the $\Gamma$ point are summarized in Table \ref{tab:irreps} for both polymorphs.

In the calculated Raman spectrum of the orthorhombic structure, as shown in Fig. \ref{fig:str_phon_Ram_ortho} (d), a dominant peak appears near 789 cm$^{-1}$ in the Lorentzian-broadened curve, accompanied by a tiny shoulder at 818 cm$^{-1}$. 
This primary peak originates from asymmetric \ce{MoO4} stretching modes, in which two oxygen atoms move inward while the other two move outward, with the central Mo atom exhibiting a slight displacement. (See Fig. \ref{fig:str_phon_Ram_ortho} (d)). 
In contrast, the symmetric \ce{MoO4} stretching mode gives rise to a weaker Raman feature at 972 cm$^{-1}$, where the Mo atoms remain nearly immobile, as discussed above. 
Only those with larger values in the results of the calculated relative Raman intensity were selected and summarized in Table \ref{tab:irreps_raman}, alongside the corresponding frequencies and irreps.
Notably, the calculated Raman spectrum is in excellent agreement with the experimentally observed Raman shifts. \cite{schumacher_unraveling_2023}

As discussed earlier, the monoclinic and orthorhombic polymorphs share nearly identical crystal structures and phonon dispersions. This structural similarity is also reflected in their Raman responses, where the calculated intensities for the monoclinic phase closely resemble those of the orthorhombic phase. See Fig. \ref{fig:str_phon_Ram_mono} (d).
This is why in the experiments, the major Raman peak at around 782 cm$^{-1}$ appears to be the same between the two temperatures without a shift of frequencies before conversion began to increase significantly.\cite{schumacher_unraveling_2023}

\begin{table}[b]
    \centering
    \caption{
    Selected phonon frequencies at $\vec{q}$ = 0 and corresponding irreducible representations that show a large value of the calculated Raman intensity ($I$), normalized to the maximum value ($I_{max}$), in orthorhombic \fmo. Those normal modes dominantly contribute to the calculated Raman intensity, as shown in Fig. \ref{fig:str_phon_Ram_ortho} (d).
    }
    \vspace{1.0 mm}
    \begin{tabular}{ccc}
    \hline\hline 
    Freq. (cm$^{-1}$) & ~~~irreps~~~ & ~~~$I$/$I_{max}$~~~ \\
    \hline
     972.05  &  $A_g$     &  0.852  \\
     819.96  &  $A_g$     &  0.150  \\
     790.28  &  $B_{1g}$  &  0.628  \\
     789.68  &  $A_g$     &  1.000  \\
     788.93  &  $B_{2g}$  &  0.216  \\
     786.87  &  $A_g$     &  0.515  \\
     786.18  &  $B_{3g}$  &  0.964  \\
    \hline\hline    
    \end{tabular} 
    \label{tab:irreps_raman}
\end{table}

\subsection{Role of oxygen vibrations in Raman}
In general, Raman intensity increases when the derivative of the polarizability with respect to a normal-mode coordinate becomes more anisotropic. 
According to Ref. \cite{raman_dft}, the Raman scattering activity can be expressed in terms of the mean polarizability derivative and the anisotropy of the polarizability tensor derivative. 
In practical DFT calculations, these quantities are obtained by evaluating the difference between two macroscopic static dielectric tensors ($\varepsilon_{ij}$) computed with respect to the phonon eigenvectors at the $\Gamma$ point $\vec{\epsilon_{i,\nu}}~(\vec{q}=0)$. 
Specifically, one dielectric tensor ($\varepsilon_+$) is calculated after a small atomic displacement along the phonon eigenvector ($\vec{r_{io}}$ + $|\vec{\epsilon_{i,\nu}}(\vec{q}=0)|$), while the other ($\varepsilon_-$) is obtained from the corresponding displacement in the opposite direction ($\vec{r_{io}}$ - $|\vec{\epsilon_{i,\nu}}(\vec{q}=0)|$). Here, $\vec{r_{io}}$ denotes the relaxed atomic position of the ith atom and $\vec{\epsilon_{i,\nu}}(\vec{q}=0)$ is the phonon eigenvector of the ith atom at frequency $\nu$. 
If the two dielectric tensors remain nearly identical in both directions, the resulting Raman scattering activity becomes correspondingly negligible, i.e., the Raman intensity is proportional to $\Delta\varepsilon$~(= $\varepsilon_+ - \varepsilon_-$ ).

According to our phonon results shown in Fig. \ref{fig:str_phon_Ram_ortho}(c), the reduction in Raman intensity around 782 cm${-1}$ is strongly correlated with oxygen-dominated phonon modes, with a minor contribution from Mo vibrations. In orthorhombic (monoclinic) \fmo, all oxygen atoms occupy the $8d$ ($4e$) Wyckoff position with $C_1$ site symmetry. Consequently, the removal of any oxygen atom eliminates all symmetry operations, reducing the structure to the triclinic $P1$ space group, in which all phonon modes become Raman active. Furthermore, our structural relaxation including an oxygen vacancy reveals significant local symmetry distortion. In general, such substantial changes in local symmetry are expected to produce frequency shifts and peak broadening in the Raman spectra \cite{kitajima_defects_1997}; however, these features are not observed in the experimental Raman measurements \cite{schumacher_unraveling_2023}, indicating that the same phonon modes are preserved. This suggests considering a supercell containing a very small concentration of defects to keep the overall local geometry effectively unchanged. However, such calculations become computationally infeasible because the required supercell would be extremely large and would lack any symmetry operations. Therefore, an alternative approach is highly necessary to effectively mimic atomic defect effects in such a large system.

\begin{figure*}[tb]
\includegraphics[width=2.0\columnwidth]{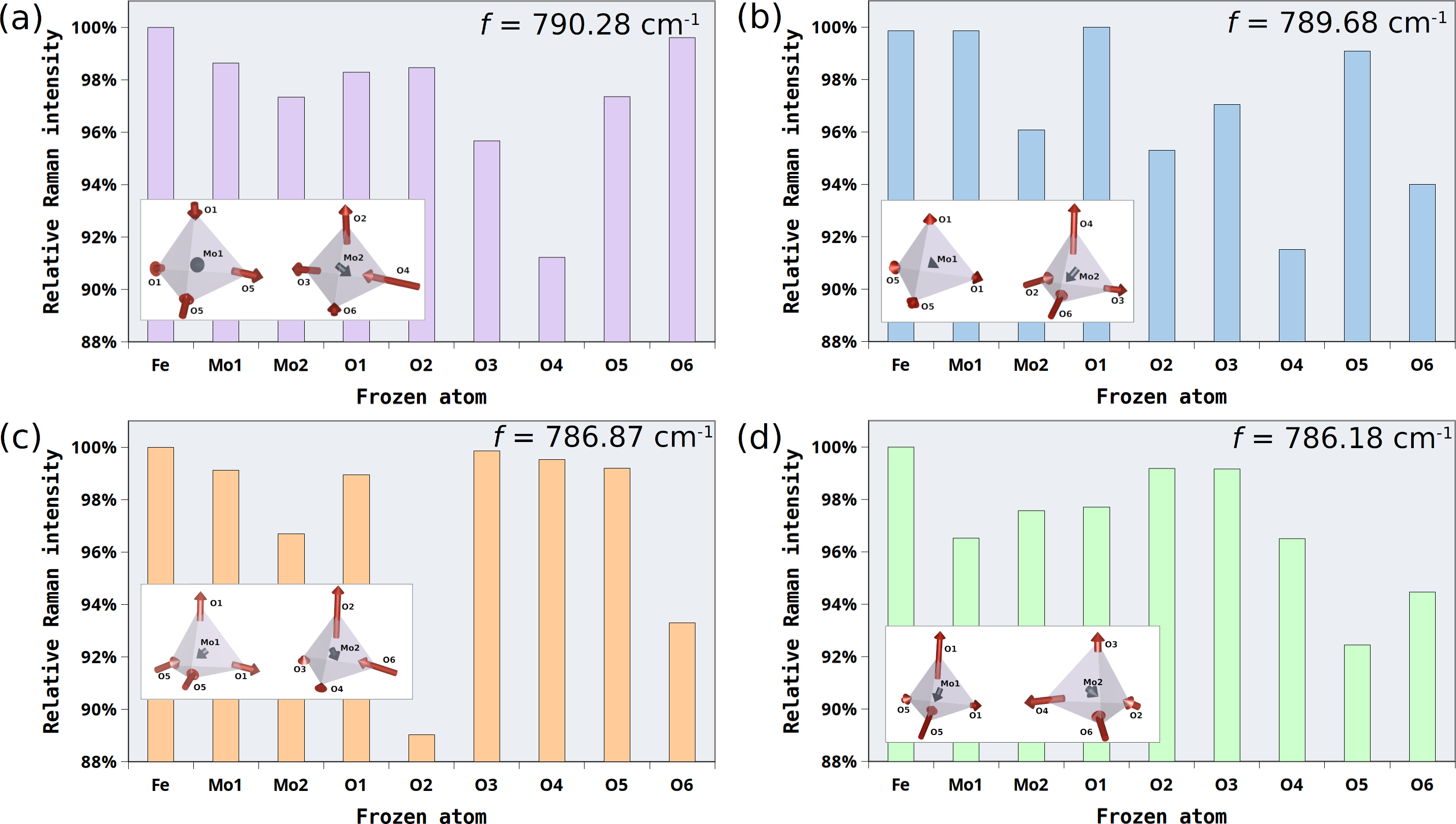} 
\caption{
Calculated relative Raman intensity at (a) 790.28 cm$^{-1}$, (b) 789.68 cm$^{-1}$, (c) 786.87 cm$^{-1}$, and (d) 786.18 cm$^{-1}$ by freezing each atom's phonon eigenvectors. In the insets, the phonon vibrations at a given frequency are visualized. Each mode is selectively frozen to analyze its contribution. (See the main text.)
Those four frequencies were selected since the calculated Raman intensity is large, contributing dominantly to the major Raman peak, as shown in Fig. \ref{fig:str_phon_Ram_ortho} (d).
}
\label{fig:frozen_raman}
\end{figure*}

To identify which atom plays a dominant role in the decrease in the observed Raman shift \cite{schumacher_unraveling_2023}, we developed an effective approach by selectively freezing specific phonon modes of the jth atom at frequency $\nu$, i.e., $\vec{\epsilon_{j,\nu}}(\vec{q}=0)$ = 0, during the Raman calculations. The resulting Raman intensity value will then differ as much as the portion of the jth atom's phonon mode contributed to the calculated Raman intensity without freezing an atom. 
Since there are 9 inequivalent atoms in orthorhombic \fmo, nine independent calculations were performed at a given frequency.
It should be noted again that this method does not account for charge effects but isolates the phonon contributions while preserving the crystal symmetry. 
As summarized in Table \ref{tab:irreps_raman}, there are four phonon modes consisting of the major peak in the calculated Raman intensity at around 789 cm$^{-1}$ in orthorhombic \fmo. See Fig. \ref{fig:str_phon_Ram_ortho} (d).
The largest Raman intensity is calculated at 789.68 cm$^{-1}$ belonging to the $A_g$ irrep.
At this frequency, our results reveal that freezing the O4 oxygen atom—i.e., suppressing its vibrational mode—leads to the largest reduction in Raman intensity, by 8.5\% relative to the unconstrained calculation.
This reduction is more than twice that observed when Mo vibrations were frozen. See Fig. \ref{fig:frozen_raman} (b).
Notably, the strongest reduction occurs when the phonon eigenvectors of Mo and O atoms are nearly antiparallel (angle $\sim$180\textdegree{}), corresponding to maximal changes in charge distribution. 
This observation explains why the reductions associated with the other three oxygen atoms (O2, O3, and O6), which also participate in the asymmetric \ce{MoO4} stretching modes (Fig. \ref{fig:str_phon_Ram_ortho} (d)), remain relatively small despite comparable atomic displacements.
Similarly, the O5 oxygen atom plays a dominant role in a decrease in the calculated Raman intensity by 7.6 \% at 786.16 cm$^{-1}$ ($B_{3g}$, where the second largest Raman intensity ($I/I_{max}$ = 0.964) is calculated. (Fig. \ref{fig:frozen_raman} (d))
That freezing the oxygen phonon vibrations results in reduced Raman intensity is also found at the remaining two frequencies, 790.28 ($B_{1g}$, $I/I_{max}$ = 0.628) and 786.87 ($A_g$, $I/I_{max}$ = 0.515) cm$^{-1}$.
In addition to the analysis of the phononic structure, these results suggest that the absence of oxygen-dominated phonon modes, which may reflect oxygen-vacancy formation, is the primary cause of the experimentally observed reduction in Raman intensity during the catalytic process in iron molybdate.

\section{Discussion and conclusion}
In the previous section, our results suggest that oxygen-related phonon modes play a dominant role in the reduction of Raman intensity. This was confirmed using an effective approach in which phonon vibrations of targeted atoms were selectively frozen while keeping the crystal structure unchanged. In this scheme, the variation of the polarizability with respect to phonon vibrations vanishes, potentially reflecting the formation of atomic vacancies. Note that this approach does not account for the distortion of the electronic cloud associated with actual atomic vacancies relative to the pristine lattice.
As mentioned above, atomic vacancies typically induce red (or blue) shifts and peak broadening in Raman spectra due to the emergence of locally disordered phases. \cite{kitajima_defects_1997} However, during the catalytic process in \fmo, no such frequency shifts or broadening were experimentally observed; instead, only a reduction in Raman intensity was reported. \cite{schumacher_unraveling_2023}
This suggests that oxygen vacancies that emerged under catalytic processes do not introduce significant structural distortions. In other words, oxygen migration may be sufficiently fast that vacant sites are rapidly reoccupied, leaving the local structure unchanged. In this scenario, the formation of oxygen vacancies should be treated dynamically rather than as static defects, which will be the subject of a future study. 

In the high-temperature orthorhombic phase of \fmo, the unit cell contains six inequivalent oxygen atoms, each forming an Fe–O–Mo bond. Removing one oxygen atom from the unit cell converts the local coordination environments from tetragonal \ce{MoO4} and octahedral \ce{FeO6} to trigonal \ce{MoO3} and pentagonal \ce{FeO5}, respectively, thereby altering the oxidation states of the associated Fe and Mo atoms. Structural relaxation of the system with a single oxygen (O4) vacancy reveals that while the trigonal \ce{MoO3} unit tends to recover its tetragonal \ce{MoO4} geometry, the pentagonal \ce{FeO5} unit remain stable, introducing defect states near the Fermi level. 
This occurs because the trigonal \ce{MoO3} unit shifts toward a neighboring \ce{FeO6} octahedron, forming an edge-sharing \ce{MoO4} tetrahedron and \ce{FeO6} octahedron. The relaxed lattice constants are obtained as $a$ = 11.778 \AA, $b$ = 8.598 \AA, and $c$ = 9.177 \AA, with angles $\alpha$ = 89.04\textdegree{}, $\beta$ = 89.07\textdegree{}, and $\gamma$ = 87.22\textdegree{}, corresponding to the triclinic $P1$ space group. Phonon calculations for this oxygen-deficient structure further indicate that the maximum (minimum) frequencies in the low- (high-) frequency region shift substantially to 486 cm$^{-1}$ (758 cm$^{-1}$), compared to 403 cm$^{-1}$ (780 cm$^{-1}$) in the stoichiometric structure. (See Fig. \ref{fig:vac_pdos_raman}) Consequently, the relaxed structure containing an oxygen vacancy deviates significantly from the stoichiometric lattice, leading to large modifications in the Raman intensity—contrary to the experimental observations. In addition, we performed Raman calculations for an unrelaxed structure containing an oxygen vacancy (O4) to keep the positions of the remaining atoms unchanged. In this case, we also observed both frequency shifts and peak broadening in the calculated Raman spectrum. (See Fig. \ref{fig:vac_pdos_raman})  
The accompanying changes in oxidation states of Fe and Mo could be directly probed using, for instance, M\"ossbauer spectroscopy. 
Therefore, these facts support the validity of our effective approach based on freezing phonon modes of targeted atoms as a reasonable means to explain the observed Raman reduction in \fmo.

Summarizing, through symmetry-mode analysis of the displacive phase transition, we confirmed that the low-temperature monoclinic and high-temperature orthorhombic phases of \fmo~are structurally nearly identical, giving rise to similar phononic characteristics. At around 789 cm$^{-1}$—where the experimentally observed Raman reduction occurs \cite{schumacher_unraveling_2023}—we found that oxygen-dominated phonons, with a minor Mo contribution, form asymmetric \ce{MoO4} stretching modes. These modes give rise to the major peak in the calculated Raman intensity, accompanied by a small shoulder at 819.96 cm$^{-1}$. In contrast, a weaker peak appears at 972.05 cm$^{-1}$, originating from symmetric \ce{MoO4} stretching modes in which Mo is barely moved.
To indirectly investigate the effects of a low concentration of atomic vacancies while preserving the symmetry of the system, we introduced an effective approach in which phonons of targeted atoms are selectively frozen.
This approach allowed us to identify that oxygen vibrations play the dominant role in the reduction of the calculated Raman intensity, suggesting that oxygen-vacancy formation may be responsible for the loss of these vibrational contributions. At the same time, the absence of measurable peak shifts or broadening in the experimental Raman spectra indicates that these vacancies do not significantly alter the local symmetry.

Structural relaxation of orthorhombic \fmo~containing a single oxygen vacancy reveals notable changes in the local coordination environment: an octahedral \ce{FeO6} unit transforms into a pentagonal \ce{FeO5} motif, accompanied by modified Fe oxidation states. Such changes could be experimentally verified using M\"ossbauer spectroscopy. However, the experimental Raman spectra suggest that, in practice, oxygen diffusion from the bulk to the surface may occur rapidly enough to effectively preserve the local symmetry.

Overall, these results suggest oxygen vacancies as the most likely origin of the reduced Raman intensity, while indicating that their presence must not significantly distort the local symmetry in order to remain consistent with experimental observations.

\section*{Acknowledgments}
We thank Christian Hess and Jan Welzenbach for helpful discussions. 
This work was funded by the Deutsche Forschungsgemeinschaft (DFG, German Research Foundation) -- CRC 1487, ``Iron, upgraded!'' -- project number 443703006.
The authors gratefully acknowledge the computing time provided to them at the NHR Center NHR@SW at Goethe University Frankfurt (project title: Computational Materials Science for Correlated systems). This is funded by the Federal Ministry of Education and Research, and the state governments participating on the basis of the resolutions of the GWK for national high performance computing at universities (\url{www.nhr-verein.de/unsere-partner}). 

\section*{Data availability}
All data created during this research are openly available at \cite{dft_data}.

\bibliography{ref_arXiv}

\newpage\onecolumngrid
\appendix
\counterwithin{figure}{section}
\counterwithin{table}{section}

\clearpage
\section{Structure comparison}

\begin{table}[h]
    \centering
    \caption{
    Pairing of the atomic positions between the transformed orthorhombic to monoclinic structure and the relaxed monoclinic structure, along with differences in the atomic positions where u$_x$, u$_y$, and u$_z$ are in relative units.
    The transformed structure from the relaxed orthorhombic structure as compiled in Table \ref{tab:str_info_ortho}, to the reference monoclinic structure,  can be obtained with the transformation matrix (See the main text).     
    }
    \vspace{1.0 mm}
    \begin{tabular}{cccccccccccccccc}
    \hline\hline \\
    \vspace{-6.5 mm} \\
    \multicolumn{5}{c}{ transformed structure (ortho. $\rightarrow$ mono.) }& ~~~~~~~ & \multicolumn{5}{c}{ relaxed monoclinic structure }& ~~~~~~~ & \\
    SG & $a$ (\AA) & $b$ (\AA) & $c$ (\AA) & $\beta$ (\textdegree{}) & &  SG & $a$ (\AA) & $b$ (\AA) & $c$ (\AA) & $\beta$ (\textdegree{}) \\
    \hline 
    $P2_1/c$ & 15.869 & 9.373 & 15.869  & 108.36 & & $P2_1/c$ & 15.869 & 9.373 & 15.868  & 108.36 & & \\
    & & & & &&&&&&&&\multicolumn{4}{c}{ differences in the atomic positions }\\
    Atom & WP  &  $x$ & $y$ & $z$ & & Atom & WP & $x$ & $y$ & $z$ & & u$_x$ & u$_y$ & u$_z$ & $|$u$|$ (\AA) \\
    \hline
    Fe1\_2  &  4e  &  0.315618  &  0.464368  &  0.935004  & &  Fe1  &  4e  &  0.315640  &  0.464360  &  0.935020  & &  0.0000  &  0.0000  &  0.0000  &  0.0004  \\
    Fe1\_3  &  4e  &  0.064996  &  0.964368  &  0.684382  & &  Fe2  &  4e  &  0.064920  &  0.964310  &  0.684360  & &  0.0001  &  0.0001  &  0.0000  &  0.0013  \\
    Fe1  &  4e  &  0.184382  &  0.964368  &  0.064996  & &  Fe3  &  4e  &  0.184400  &  0.964400  &  0.065010  & &  0.0000  &  0.0000  &  0.0000  &  0.0004  \\
    Fe1\_4  &  4e  &  0.435004  &  0.464368  &  0.315618  & &  Fe4  &  4e  &  0.434910  &  0.464460  &  0.315590  & &  0.0001  &  -0.0001  &  0.0000  &  0.0017  \\
    Mo2  &  4e  &  0.488266  &  0.750000  &  0.488266  & &  Mo1  &  4e  &  0.488270  &  0.749990  &  0.488300  & &  0.0000  &  0.0000  &  0.0000  &  0.0005  \\
    Mo3\_2  &  4e  &  0.126709  &  0.621198  &  0.771860  & &  Mo2  &  4e  &  0.126750  &  0.621220  &  0.771880  & &  0.0000  &  0.0000  &  0.0000  &  0.0007  \\
    Mo3\_4  &  4e  &  0.271861  &  0.621198  &  0.126709  & &  Mo3  &  4e  &  0.271760  &  0.621150  &  0.126630  & &  0.0001  &  0.0000  &  0.0001  &  0.0018  \\
    Mo3  &  4e  &  0.373291  &  0.121198  &  0.228139  & &  Mo4  &  4e  &  0.373370  &  0.121190  &  0.228190  & &  -0.0001  &  0.0000  &  -0.0001  &  0.0013  \\
    Mo3\_3  &  4e  &  0.228139  &  0.121198  &  0.873291  & &  Mo5  &  4e  &  0.228070  &  0.121230  &  0.873230  & &  0.0001  &  0.0000  &  0.0001  &  0.0012  \\
    Mo2\_2  &  4e  &  0.011734  &  0.250000  &  0.011734  & &  Mo6  &  4e  &  0.011760  &  0.250000  &  0.011730  & &  0.0000  &  0.0000  &  0.0000  &  0.0004  \\
    O4\_2  &  4e  &  0.963756  &  0.880951  &  0.404355  & &  O1  &  4e  &  0.963940  &  0.881170  &  0.404490  & &  -0.0002  &  -0.0002  &  -0.0001  &  0.0037  \\
    O6\_3  &  4e  &  0.149166  &  0.921168  &  0.173009  & &  O2  &  4e  &  0.149280  &  0.921110  &  0.173070  & &  -0.0001  &  0.0001  &  -0.0001  &  0.0018  \\
    O5\_2  &  4e  &  0.096187  &  0.697068  &  0.269305  & &  O3  &  4e  &  0.096210  &  0.697050  &  0.269320  & &  0.0000  &  0.0000  &  0.0000  &  0.0004  \\
    O7\_3  &  4e  &  0.040258  &  0.989272  &  0.282000  & &  O4  &  4e  &  0.040300  &  0.989290  &  0.281980  & &  0.0000  &  0.0000  &  0.0000  &  0.0008  \\
    O6  &  4e  &  0.173009  &  0.921168  &  0.649166  & &  O5  &  4e  &  0.172890  &  0.921150  &  0.649050  & &  0.0001  &  0.0000  &  0.0001  &  0.0022  \\
    O7\_2  &  4e  &  0.282000  &  0.989272  &  0.540258  & &  O6  &  4e  &  0.281950  &  0.989440  &  0.540250  & &  0.0001  &  -0.0002  &  0.0000  &  0.0017  \\
    O4\_3  &  4e  &  0.404355  &  0.619049  &  0.963756  & &  O7  &  4e  &  0.404410  &  0.618980  &  0.963890  & &  -0.0001  &  0.0001  &  -0.0001  &  0.0021  \\
    O5\_3  &  4e  &  0.269305  &  0.802932  &  0.096187  & &  O8  &  4e  &  0.269230  &  0.802850  &  0.096030  & &  0.0001  &  0.0001  &  0.0002  &  0.0025  \\
    O4  &  4e  &  0.463756  &  0.880951  &  0.904355  & &  O9  &  4e  &  0.463660  &  0.880860  &  0.904330  & &  0.0001  &  0.0001  &  0.0000  &  0.0017  \\
    O8\_2  &  4e  &  0.995950  &  0.828949  &  0.591502  & &  O10  &  4e  &  0.995890  &  0.828700  &  0.591570  & &  0.0001  &  0.0002  &  -0.0001  &  0.0029  \\
    O8\_4  &  4e  &  0.091502  &  0.828949  &  0.995950  & &  O11  &  4e  &  0.091420  &  0.829180  &  0.995980  & &  0.0001  &  -0.0002  &  0.0000  &  0.0026  \\
    O8  &  4e  &  0.495950  &  0.828949  &  0.091502  & &  O12  &  4e  &  0.495990  &  0.829070  &  0.091520  & &  0.0000  &  -0.0001  &  0.0000  &  0.0013  \\
    O9  &  4e  &  0.223370  &  0.905468  &  0.363325  & &  O13  &  4e  &  0.223400  &  0.905400  &  0.363380  & &  0.0000  &  0.0001  &  -0.0001  &  0.0011  \\
    O7\_4  &  4e  &  0.540258  &  0.510728  &  0.282000  & &  O14  &  4e  &  0.540110  &  0.510860  &  0.281910  & &  0.0001  &  -0.0001  &  0.0001  &  0.0026  \\
    O9\_4  &  4e  &  0.363325  &  0.594532  &  0.223370  & &  O15  &  4e  &  0.363200  &  0.594580  &  0.223340  & &  0.0001  &  0.0000  &  0.0000  &  0.0019  \\
    O6\_4  &  4e  &  0.350834  &  0.421168  &  0.826991  & &  O16  &  4e  &  0.350900  &  0.421250  &  0.827030  & &  -0.0001  &  -0.0001  &  0.0000  &  0.0013  \\
    O7  &  4e  &  0.218000  &  0.489272  &  0.459742  & &  O17  &  4e  &  0.217910  &  0.489150  &  0.459710  & &  0.0001  &  0.0001  &  0.0000  &  0.0018  \\
    O9\_2  &  4e  &  0.276630  &  0.405468  &  0.636675  & &  O18  &  4e  &  0.276750  &  0.405610  &  0.636690  & &  -0.0001  &  -0.0001  &  0.0000  &  0.0023  \\
    O6\_2  &  4e  &  0.326991  &  0.421168  &  0.350834  & &  O19  &  4e  &  0.326920  &  0.421230  &  0.350810  & &  0.0001  &  -0.0001  &  0.0000  &  0.0012  \\
    O8\_3  &  4e  &  0.591502  &  0.828949  &  0.495950  & &  O20  &  4e  &  0.591530  &  0.828840  &  0.495970  & &  0.0000  &  0.0001  &  0.0000  &  0.0011  \\
    O5  &  4e  &  0.403813  &  0.302932  &  0.230695  & &  O21  &  4e  &  0.403770  &  0.302970  &  0.230720  & &  0.0000  &  0.0000  &  0.0000  &  0.0010  \\
    O4\_4  &  4e  &  0.095645  &  0.119049  &  0.036244  & &  O22  &  4e  &  0.095780  &  0.119250  &  0.036460  & &  -0.0001  &  -0.0002  &  -0.0002  &  0.0039  \\
    O9\_3  &  4e  &  0.136675  &  0.094532  &  0.776630  & &  O23  &  4e  &  0.136610  &  0.094560  &  0.776550  & &  0.0001  &  0.0000  &  0.0001  &  0.0014  \\
    O5\_4  &  4e  &  0.230695  &  0.302932  &  0.903813  & &  O24  &  4e  &  0.230670  &  0.302990  &  0.903700  & &  0.0000  &  -0.0001  &  0.0001  &  0.0018  \\
    \hline\hline    
    \end{tabular} 
    \label{tab:str_info_diff}
\end{table}

\clearpage
\section{Phonon in monoclinic \fmo}

\begin{figure}[h]
\includegraphics[width=1.0\columnwidth]{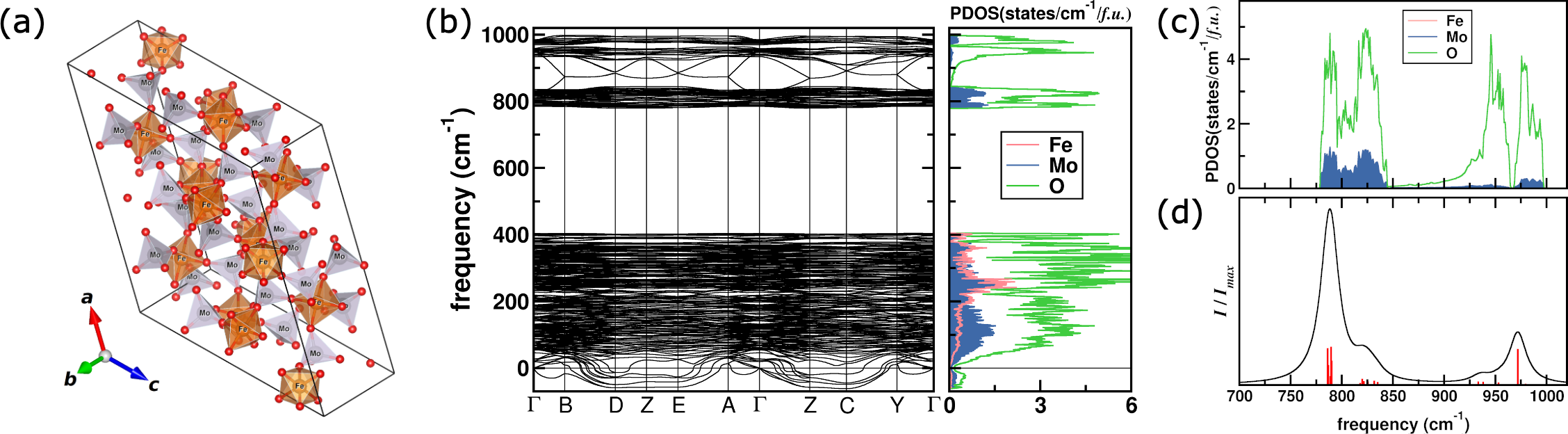} 
\caption{
(a) Crystal structure of low-temperature monoclinic \fmo. (b) The corresponding phonon dispersions along the high symmetry points and atom-resolved phonon density of states. (c) Enlarged atom-resolved phonon density of states of the high-frequency regime. (d) Calculated Raman intensity ($I$), normalized to the maximum value ($I_{max}$), depicted with a thin red bar, as a function of frequency. 
Lorentzian broadening with a width of 10 cm$^{-1}$ is plotted with a black solid line. 
At 789 cm$^{-1}$, asymmetric \ce{MoO4} vibrational modes are dominant, while symmetric ones are found at 971 cm$^{-1}$.
A red (black) arrow represents the O(Mo)-atom vibrations at a given frequency.  
}
\label{fig:str_phon_Ram_mono}
\end{figure}

As we discussed in the main text, both orthorhombic and monoclinic \fmo~share the almost identical crystal structure, demonstrated by a symmetry-mode analysis of the displacive phase transition. Figure \ref{fig:str_phon_Ram_mono} illustrates the crystal structure, phonon dispersions, atom-resolved phonon density of states, and the corresponding Raman intensity in of monoclinic \fmo. Comparing this figure with Fig. \ref{fig:str_phon_Ram_ortho} shows how the phononic structures and the corresponding Raman intensity are similar to each other. 
It should be noted that the presence of the artificial imaginary frequencies arises from the finite unit cell size effect, which should disappear by adopting a supercell, as confirmed in the orthorhombic phase. (See the main text.)

\clearpage
\section{}
We summarized all Raman-related optical phonon frequencies at $\vec{q}$ = 0 with their irreps, and the corresponding normalized Raman intensity ($I$/$I_{max}$) in monoclinic and orthorhombic \fmo.

\begin{table}[h]
    \centering
    \caption{ 
    Optical phonon frequencies at $\vec{q}$ = 0 with their irreps that are responsible for Raman processes, and the corresponding calculated Raman intensity ($I$), normalized to the maximum value ($I_{max}$), in monoclinic and orthorhombic \fmo. }
    \vspace{1.0 mm}
    \begin{tabular}{ccccccccccc}
    \hline\hline 
    \multicolumn{7}{c}{Monoclinic} &   & \multicolumn{3}{c}{Orthorhombic}   \\
    \vspace{-2.5 mm} \\
    Freq. (cm$^{-1}$) & Irreps & ~~~$I$/$I_{max}$~~~ & ~~~  & Freq. (cm$^{-1}$) & Irreps & ~~~$I$/$I_{max}$~~~ & ~~~~~~~~~~~~~~~~~  & Freq. (cm$^{-1}$) & Irreps & ~~~$I$/$I_{max}$~~~ \\
    \hline
     979.71   &   $B_g$   &   0.006   &   &   825.53   &   $B_g$   &   0.074   &      &   979.72   &   $B_{3g}$   &   0.017   \\
     979.69   &   $A_g$   &   0.001   &   &   824.03   &   $A_g$   &   0.000   &      &   977.86   &   $B_{1g}$   &   0.001   \\
     979.61   &   $B_g$   &   0.014   &   &   823.97   &   $B_g$   &   0.000   &      &   972.71   &   $B_{2g}$   &   0.000   \\
     977.77   &   $A_g$   &   0.001   &   &   821.69   &   $B_g$   &   0.079   &      &   972.07   &   $A_g$   &   0.852   \\
     974.85   &   $A_g$   &   0.000   &   &   821.45   &   $A_g$   &   0.000   &      &   956.52   &   $B_{2g}$   &   0.006   \\
     974.81   &   $B_g$   &   0.001   &   &   821.43   &   $B_g$   &   0.001   &      &   953.12   &   $A_g$   &   0.046   \\
     972.64   &   $B_g$   &   0.000   &   &   819.98   &   $A_g$   &   0.145   &      &   950.85   &   $B_{1g}$   &   0.002   \\
     971.98   &   $A_g$   &   0.942   &   &   817.96   &   $A_g$   &   0.035   &      &   938.20   &   $B_{2g}$   &   0.062   \\
     956.50   &   $B_g$   &   0.006   &   &   816.54   &   $B_g$   &   0.001   &      &   933.28   &   $B_{3g}$   &   0.068   \\
     953.74   &   $A_g$   &   0.000   &   &   809.69   &   $A_g$   &   0.000   &      &   835.25   &   $A_g$   &   0.037   \\
     953.66   &   $B_g$   &   0.001   &   &   809.63   &   $B_g$   &   0.002   &      &   831.91   &   $B_{3g}$   &   0.096   \\
     953.12   &   $A_g$   &   0.044   &   &   809.51   &   $B_g$   &   0.001   &      &   825.68   &   $B_{1g}$   &   0.078   \\
     950.86   &   $A_g$   &   0.001   &   &   806.84   &   $B_g$   &   0.000   &      &   825.67   &   $B_{3g}$   &   0.083   \\
     949.99   &   $A_g$   &   0.000   &   &   806.81   &   $A_g$   &   0.000   &      &   821.68   &   $B_{2g}$   &   0.079   \\
     949.98   &   $B_g$   &   0.000   &   &   803.64   &   $A_g$   &   0.004   &      &   819.96   &   $A_g$   &   0.150   \\
     946.27   &   $B_g$   &   0.000   &   &   793.27   &   $B_g$   &   0.000   &      &   817.98   &   $B_{1g}$   &   0.035   \\
     946.24   &   $A_g$   &   0.000   &   &   793.26   &   $A_g$   &   0.000   &      &   816.62   &   $B_{2g}$   &   0.001   \\
     938.13   &   $B_g$   &   0.062   &   &   790.25   &   $A_g$   &   0.629   &      &   809.51   &   $B_{3g}$   &   0.003   \\
     933.34   &   $B_g$   &   0.075   &   &   789.68   &   $A_g$   &   1.000   &      &   803.46   &   $B_{1g}$   &   0.003   \\
     835.08   &   $A_g$   &   0.061   &   &   788.99   &   $B_g$   &   0.217   &      &   790.27   &   $B_{1g}$   &   0.628   \\
     831.75   &   $B_g$   &   0.096   &   &   786.96   &   $A_g$   &   0.515   &      &   789.68   &   $A_g$   &   1.000   \\
     828.62   &   $A_g$   &   0.000   &   &   786.28   &   $B_g$   &   0.964   &      &   788.92   &   $B_{2g}$   &   0.216   \\
     828.56   &   $B_g$   &   0.000   &   &   784.38   &   $A_g$   &   0.000   &      &   786.86   &   $A_g$   &   0.515   \\
     825.62   &   $A_g$   &   0.074   &   &   784.33   &   $B_g$   &   0.000   &      &   786.18   &   $B_{3g}$   &   0.964   \\      
    \hline\hline    
    \end{tabular} 
    \label{tab:irreps}
\end{table}

\clearpage
\section{Phonon DOS and Raman intensity including O vacancy}
\begin{figure}[h]
\includegraphics[width=0.6\columnwidth]{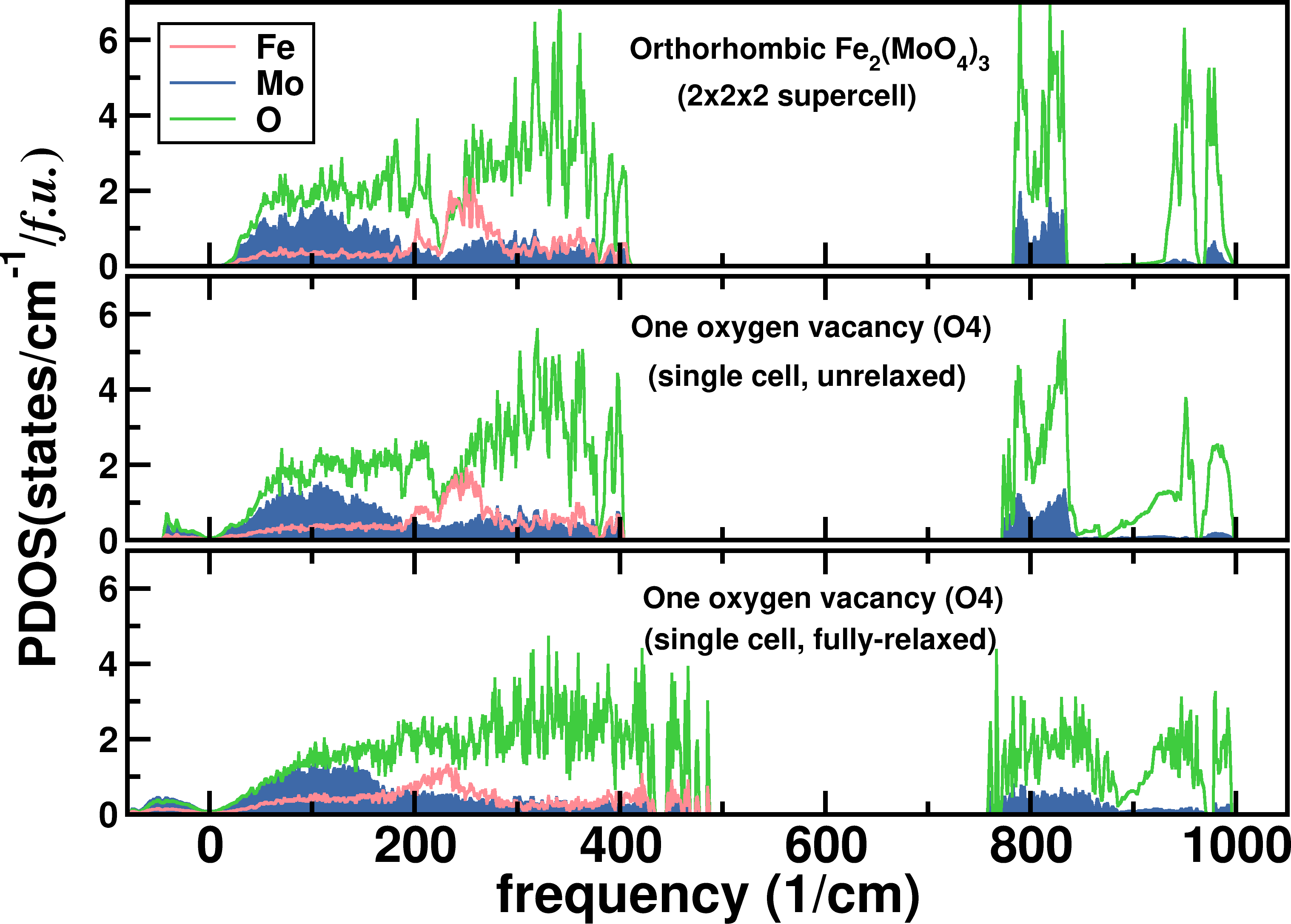} 
\hspace{1cm}
\includegraphics[width=0.3\columnwidth]{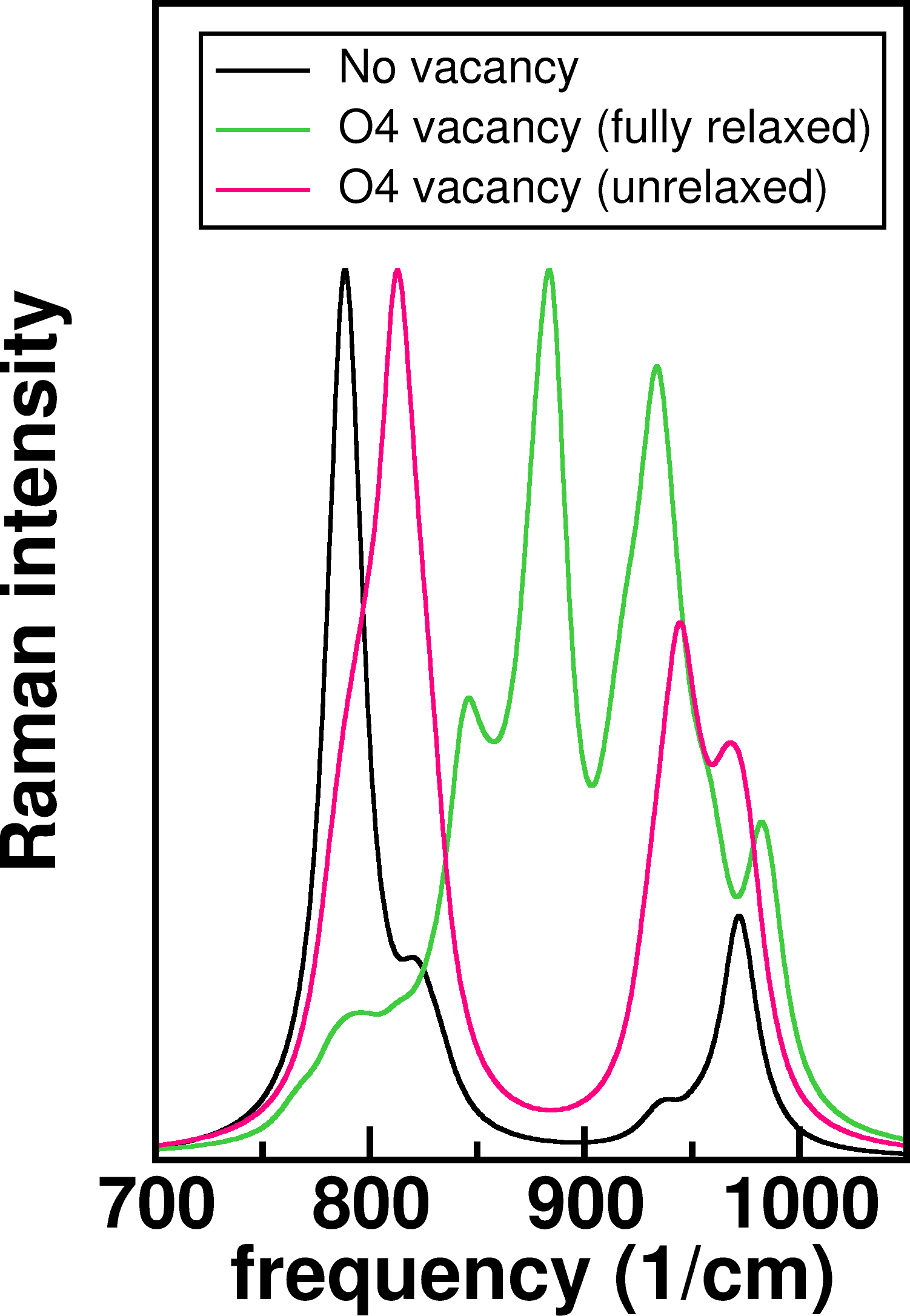} 
\caption{
(Left) Comparison of the atom-resolved phonon density of states in orthorhombic \fmo~without (top) and with (middle, bottom) an oxygen vacancy. (Right) The corresponding Raman intensity results, where the maximum heights in each case are aligned equally for comparison.
}
\label{fig:vac_pdos_raman}
\end{figure}

\end{document}